\documentclass[a4paper]{article}

\usepackage{INTERSPEECH2022}
\usepackage{enumitem}
\usepackage{graphicx}
\usepackage{booktabs}
\usepackage{multirow}
\usepackage{colortbl}
\usepackage{amssymb}
\usepackage{amsmath}
\usepackage{pifont}
\usepackage{cite}
\newcommand{\cmark}{\ding{51}}%
\newcommand{\xmark}{\ding{55}}%
\definecolor{mygray}{gray}{.9}

\title{Beam-Guided TasNet: An Iterative Speech Separation Framework with Multi-Channel Output}
\name{Hangting~Chen$^{1,2}$, Yang~Yi$^{1,2}$,  Dang~Feng$^{1,2}$ and Pengyuan~Zhang$^{1,2}$
\thanks{This work is partially supported by the National Natural Science Foundation of China (Nos.62071461).
}\thanks{Pengyuan Zhang is the corresponding author.}
}
\address{
	$^1$Key Laboratory of Speech Acoustics \& Content Understanding, Institute of Acoustics, CAS, China\\
	$^2$University of Chinese Academy of Sciences, Beijing, China
}
\email{\{chenhangting,yangyi,dangfeng,zhangpengyuan\}@hccl.ioa.ac.cn}

\begin{document}

\maketitle
%
\begin{abstract}
Time-domain audio separation network (TasNet) has achieved remarkable performance in blind source separation (BSS). Classic multi-channel speech processing framework employs signal estimation and beamforming. For example, Beam-TasNet links multi-channel convolutional TasNet (MC-Conv-TasNet) with minimum variance distortionless response (MVDR) beamforming, which leverages the strong modeling ability of data-driven network and boosts the performance of beamforming with an accurate estimation of speech statistics. Such integration can be viewed as a directed acyclic graph by accepting multi-channel input and generating multi-source output. In this paper, we design a ``multi-channel input, multi-channel multi-source output'' (MIMMO) speech separation system entitled ``Beam-Guided TasNet'', where MC-Conv-TasNet and MVDR can interact and promote each other more compactly under a directed cyclic flow. Specifically, the first stage uses Beam-TasNet to generate estimated single-speaker signals, which favors the separation in the second stage. The proposed framework facilitates iterative signal refinement with the guide of beamforming and seeks to reach the upper bound of the MVDR-based methods. Experimental results on the spatialized WSJ0-2MIX demonstrate that the Beam-Guided TasNet has achieved an SDR of 21.5 dB, exceeding the baseline Beam-TasNet by 4.1 dB under the same model size and narrowing the gap with the oracle signal-based MVDR to 2 dB.
\end{abstract}
\noindent\textbf{Index Terms}: Speech separation, multi-channel speech processing, MVDR, time-domain network

\section{Introduction}

Speech separation has achieved remarkable advances since the introduction of deep learning. When a microphone array captures a speech signal, spatial information can be leveraged to separate sources from different directions. A conventional framework consists of mask estimation, beamforming, and an optional post-filtering for ``multi-channel input, multi-source output'' \cite{Erdogan2016ImprovedMB,Kanda2019GuidedSS}. The minimum variance distortionless response (MVDR) beamformer requires estimation of the spatial correlation matrices (SCMs), typically computed based on the estimated speech and noise masks. Since the considerable speech separation performance achieved by the time-domain audio separation network (TasNet) \cite{Luo2019ConvTasNetSI}, the recently proposed Beam-TasNet \cite{Ochiai2020BeamTasNetTA} uses the estimated time-domain signals to compute the SCMs, which has outperformed the MVDR based on the oracle frequency-domain masks.


In this paper, we adopt ``multi-channel input, multi-channel multi-source output" (MIMMO) for the first time to design a multi-channel separation framework entitled ``Beam-Guided TasNet'', which shows a promising potential of learning data-driven models guided by beamforming. Specifically, the framework utilizes two sequential Beam-TasNets for 2-stage processing. The first stage uses a multi-channel convolutional TasNet (MC-Conv-TasNet) and the MVDR beamforming to perform blind source separation (BSS). In the second stage, an MC-Conv-TasNet guided by MVDR-beamformed signals can refine separated signals iteratively. Experiments on the spatialized WSJ0-2MIX \cite{Wang2018MultiChannelDC} exhibited significant performance improvement compared with the baseline Beam-TasNet.\footnote{The core codes are available at \\ https://github.com/hangtingchen/Beam-Guided-TasNet} The contributions are as follows:

1. The directed cyclic flow of the second stage promotes the MC-Conv-TasNet and MVDR iteratively and seeks to reach the upper bound of the MVDR-based methods, which obtained an SDR of $19.1$ dB.

2. The unfolding training further improves the performance to $21.5$ dB, which narrowed the gap between the estimated and oracle signal-based MVDR to $2$ dB.

3. A causal Beam-Guide TasNet is explored for online processing, illustrating that the Beam-Guided TasNet is effective even though the utterance-level information is unreachable. The performance degradation caused by causality was alleviated, with SDRs improved from $11.4$ dB to $14.0$ dB by replacing Beam-TasNet with the Beam-Guided TasNet.

\section{The proposed Beam-Guided TasNet}

\begin{figure*}[thp]
	\centering
	\includegraphics[width=0.90\linewidth]{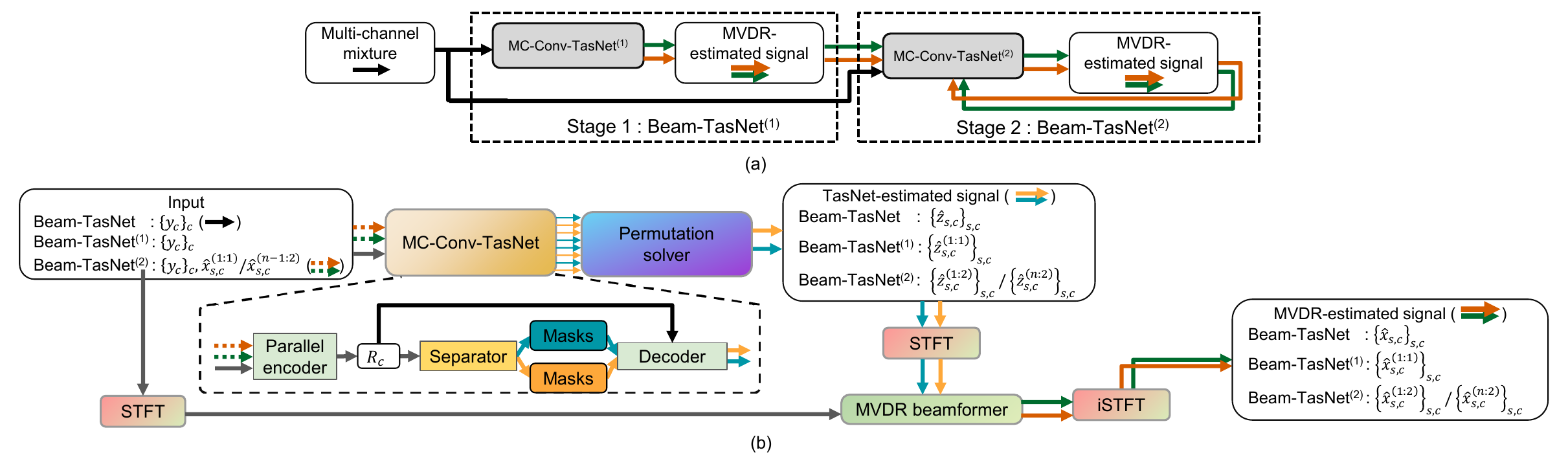}
	\caption{(a) Beam-Guided TasNet with a 2-stage framework for iterative refinement. (b) The signal processing routine in the Beam-TasNet, the first and the second stage model. The dashed lines are the additional input for the second stage model.}
	\label{fig:framework}
\end{figure*}

\subsection{Beam-TasNet}

Suppose that speech signals from $S$ sources are captured by $C$ microphones,
\vspace{-0.2cm}
\begin{equation}
y_c=\sum_{s=1}^{S}x_{s,c}.
\end{equation}
The Beam-TasNet integrates the time-domain network and the beamforming to estimate signal image $x_{s,c}$ on microphone $c$ from source $s$ with a given mixture $y_{c}$. As plotted in Fig.\ref{fig:framework}(b), the baseline Beam-TasNet is mainly composed of an MC-Conv-TasNet \cite{Gu2019EndtoEndMS}, a permutation solver, and an MVDR beamformer. Given a multi-channel input ${\{y_c\}}_c$ indicating a collection of $y_c$ along channels ($c=1,...,C$), MC-Conv-TasNet generates $\hat{x}_{s,c}$ representing the estimated image of source $s$ on channel $c$. The MC-Conv-TasNet utilizes a parallel encoder ($\text{ParEnc}$) for encoding the input multi-channel signal into a 2-dimensional temporal-spectro representation $R_c$ \cite{Gu2019EndtoEndMS}:
\begin{equation}
\label{eq:rc}
	R_c=\text{ParEnc}({\{y_c\}}_c,c),
\end{equation}
a separator to estimate the temporal-spectro masks:
\begin{equation}
	{\{\hat{M}_{s,c}\}}_{s}=\text{Seperator}(R_c),
\end{equation}
and a decoder to recover the single-speaker waveform:
\begin{equation}
	\hat{z}_{s,c}=\text{Dec}(\hat{M}_{s,c}\odot R_c),
\end{equation}
where $\odot$ is Hadamard product, $c$ indicates the reference channel and can be determined by the order of the input. The permutation solver determines the source order by comparing the similarity across channels with the output of the first channel.
The MVDR beamformer accepts the reordered estimation and calculates the SCM for each source,
\begin{equation}
	\resizebox{0.5\linewidth}{!}{$
	\mathbf{\hat{\Phi}}_{f}^{\mathrm{Target}_{s}}=\frac{1}{T} \sum_{t=1}^{T} \hat{\mathbf{Z}}_{s, t, f} \hat{\mathbf{Z}}_{s, t, f}^{\mathrm{H}}
	$}
\end{equation}
\begin{equation}
	\resizebox{0.85\linewidth}{!}{$
	\mathbf{\hat{\Phi}}_{f}^{\mathrm{Interfer}_{s}}=\frac{1}{T} \sum_{t=1}^{T}(\mathbf{Y}_{t, f}-\hat{\mathbf{Z}}_{s, t, f})(\mathbf{Y}_{t, f}-\hat{\mathbf{Z}}_{s, t, f})^{\mathrm{H}}
	$}
\end{equation}
where $\mathbf{\hat{\Phi}}_{f}^{\mathrm{{Target}_{s}}}$/$\mathbf{\hat{\Phi}}_{f}^{\mathrm{{Interfer}_{s}}}$ denotes the speech/interference SCMs for source $s$, $\mathbf{Y}$ and $\mathbf{\hat{Z}}$ denotes the short-time Fourier transform (STFT) spectra of $\{y_{c}\}_{c}$ and $\{\hat{z}_{s,c}\}_{c}$, $\cdot^{H}$ denotes Hermitian transpose. The signal enhanced by the MVDR beamforming is calculated by
\begin{equation}
	\hat{x}_{s,c}=\text{MVDR}(\mathbf{\Phi}_{f}^{\mathrm{Target}_{s}},\mathbf{\Phi}_{f}^{\mathrm{Interfer}_{s}},c)^\mathrm{H}\mathbf{Y}_{t,f},
\end{equation}
where reference channel $c$ is indicated by a one-hot vector \cite{Souden2010OnOF}. 

In summary, the Beam-TasNet uses MC-Conv-TasNet to estimate SCMs $\mathbf{\hat{\Phi}}$ with the estimated multi-channel image signals $\{\hat{z}_{s,c}\}_{s,c}$ ($\text{MC-Conv-TasNet}(\mathbf{\hat{\Phi}}|y_c)$) and uses MVDR beamforming to estimate $\hat{x}_{s,c}$ ($\text{MVDR}(\hat{x}_{s,c}|y_c,\mathbf{\hat{\Phi}})$), which can be formulated as
\begin{equation}
	\{\hat{x}_{s,c}\}_{s}=\text{Beam-TasNet}(\{y_c\}_{c},c).
\end{equation}
with each channel served as the reference channel and then do beamforming on the reference channel $c$.

\subsection{MIMMO model}
MC-Conv-TasNet uses different channel orders to obtain temporal-spectro representation for the reference channel (Eq.\ref{eq:rc}), for example, $R_1$ for channel order [1,2,3,4] and  $R_4$ for channel order [4,1,2,3]. To obtain estimated signal, MC-Conv-TasNet needs to be run in $C$ times, where $C$ is the channel number. We call estimating the reference channel as ``multi-channel input, single-channel multi-source output'' (MISMO). For fast inference, we adopt MIMMO inference on MC-Conv-TasNet. The network accept ${\{y_c\}}_{c}$ to generate
\begin{equation}
\label{eq:rc2}
	R=\text{ParEnc}({\{y_c\}}_{c}),
\end{equation}
the separator estimates the temporal-spectro masks for all channels and sources:
\begin{equation}
	{\{\hat{M}_{s,c}\}}_{s,c}=\text{Seperator}(R),
\end{equation}
and the parallel decoder recovers the single-speaker waveform:
\begin{equation}
	\hat{z}_{s,c}=\text{ParDec}(\hat{M}_{s,c}\odot R),
\end{equation}
where $\text{ParDec}$ generates signals for different channels using different decoders. MC-Conv-Tasnet only need to be run in one time to get estimated signals for all sources and channels.

\subsection{Beam-Guided TasNet}

As plotted in Fig.\ref{fig:framework}(a), the first stage in the Beam-Guided TasNet employs the original Beam-TasNet, which performs BSS with the MVDR beamforming. In the second stage, the network performs source separation additionally guided by the beamformed signal. The encoder of the MC-Conv-TasNet in the second stage accepts ($C+S\times C$) channels, including $C$-channel mixtures and $S\times C$-speaker beamformed signals.

As shown in Fig.\ref{fig:framework}(b), we first feed the mixture signal $y_c$ through $\text{Beam-TasNet}^{(1)}$ to obtain the enhanced single-speaker signals $\hat{x}_{s,c}^{(1)}$, 
\begin{equation}
	\{\hat{x}_{s,c}^{(1)}\}_{s,c}=\text{Beam-TasNet}^{(1)}(\{y_c\}_{c}).
\end{equation}
Then the second stage uses a second Beam-TasNet to accept $\hat{x}_{s,c}^{(1)}$ and $y_c$ and to generate $\hat{x}_{s,c}^{(2:1)}$,
\begin{equation}
	\{\hat{x}_{s,c}^{(2:1)}\}_{s,c}=\text{Beam-TasNet}^{(2)}(\{y_c\}_{c},\{\hat{x}_{s,c}^{(1)}\}_{s,c}).
\end{equation}
where superscript $\cdot^{(2:1)}$ indicates that the signal is generated by the second stage in the first iteration. In such a way, the second Beam-TasNet integrates the strength of the MVDR beamforming into the data-driven model. Different from target speaker extraction \cite{molkov2019SpeakerBeamSA} and neural spatial filtering \cite{Gu2019NeuralSF,Gu2020TemporalSpatialNF}, we deduce the source information by the enhanced signal calculated by the MVDR beamforming. 

The framework leads to a directed cyclic flow of multi-channel signals with iterative refinement implemented on the second stage (Fig.\ref{fig:framework}(a)). MIMMO is achieved by separately setting each channels as the reference channel in the MVDR beamforming. The second stage can iteratively accept $\hat{x}_{s,c}^{(2:n-1)}$ and generate $\hat{x}_{s,c}^{(2:n)}$,
\begin{equation}
	\{\hat{x}_{s,c}^{(2:n)}\}_{s,c}=\text{Beam-TasNet}^{(2)}(\{y_c\}_{c},\{\hat{x}_{s,c}^{(2:n-1)}\}_{s,c}),
\end{equation}
where $n=2,3,...$ denotes the iteration number. In summary, the MVDR beamforming estimates the distortionless signals with the given SCMs ($\text{MVDR}(\hat{x}_{s,c}^{(2:n)}|y_c,\mathbf{\hat{\Phi}}^{(2:n-1)})$); MC-Conv-TasNet finds an optimal set of SCMs with the given distortionless signals ($\text{MC-Conv-TasNet}(\mathbf{\hat{\Phi}}^{(2:n)}|y_c,\hat{x}_{s,c}^{(2:n)})$).

In the training procedure, we unfolds the second stage for source-to-noise ratio (SNR) loss calculation to help the iterative refinement in the second stage,
\begin{equation}
	\label{eq:unfolding_loss}
	L=-\text{SNR}(\hat{z}_{s,c}^{(1)},x_{s,c})-\text{SNR}(\hat{z}_{s,c}^{(2:1)},x_{s,c})-\text{SNR}(\hat{z}_{s,c}^{(2:2)},x_{s,c}).
\end{equation}
Since MC-Conv-TasNet with MIMMO can infer all channel in one pass, we can train the whole network in an end-to-end way, \textit{i.e.}, we do not need to train different stages sequentially. 

\subsection{The causal variant}

Compared with non-causal models, the causal variant only uses the current and the past audio information, which can be deployed for online processing. A causal Beam-Guided TasNet uses the causal MC-Conv-TasNet and the frame-by-frame updated MVDR. We use channel-wise layer normalization to replace global layer normalization \cite{Luo2019ConvTasNetSI,Sonning2020PerformanceSO}. The permutation solver and MVDR are updated in a frame-by-frame way, whose formulas can be found in Appendix A\footnote{https://github.com/hangtingchen/Beam-Guided-TasNet/blob/main/INTERSPEECH\_2022\_Appendix.pdf}.

\subsection{Relation with other works}
Beam-guided separation is similar to deep unfolding (DU), extending iteration steps into network layers. The significant differences are two-fold. First, DU uses untied parameters for different iteration steps \cite{DBLP:journals/corr/HersheyRW14}. The proposed method uses shared parameters in the second stage for different iteration numbers. Second, DU combines the deep learning-based method with existing model-based methods. However, few model-based methods have studied iterating beamforming and signal estimation. A theoretical discussion can be found in Appendix B.

Some researchers have used second-stage networks but do not explore iterative refinement \cite{DBLP:journals/taslp/WangWW21}. The method in \cite{DBLP:conf/apsipa/TogamiMKYK20} conducts computer-resource-aware deep speech separation (CRA-DSS). The major differences are three-fold. First, CRA-DSS uses untied parameters for different blocks, similar to DU. Second, the proposed second stage uses both $y_c$ and MVDR generated signal, while CRA-DSS only uses MVDR signal. The important role played by $y_c$ will be stated in Section \ref{sec:res_dis}. Third, CRA-DASS trains blocks sequentially while our MIMMO and unfolding training make it can be trained end-to-end.

\section{Experimental setup}

We evaluate the proposed framework on the spatialized WSJ0-2MIX corpus \cite{Wang2018MultiChannelDC}. The reverberant mixtures were generated by convolving the room impulse responses (RIRs) with the clean single-speaker utterances. The RIRs were randomly sampled with sound decay time (T60) from 0.2s to 0.6s. The signal-to-interference ratio was sampled from $-5$ dB to $+5$ dB. The dataset contains $20,000$ ($\sim30h$), $5,000$ ($\sim10h$), and $3,000$ ($\sim5h$) multi-channel two-speaker mixtures in the training, development and evaluation sets.
Two dataset variations are available: a ``min'' version where the longer signal is truncated, and a ``max'' version where silence is appended to the shorter signal \cite{WichernAFZMCMR19}. The training and the development sets were generated with a sampling rate of 8kHz and a mode of ``min''; the testing set was generated with a sampling rate of 8kHz and a mode of ``max'' for word error rates (WERs) evaluation.

The first $4$ channels out of $8$ were used to train and evaluate the models for a fair comparison with \cite{Ochiai2020BeamTasNetTA}. In evaluation, the default first channel was chosen as the reference. The window settings of the STFT were set as a $512$ ms frame length and a $128$ ms hop size in MVDR due to the considerable reverberant time. In the frame-by-frame processing, the MVDR calculation was performed frame-wisely to obtain the SCMs, MVDR filters, and enhanced signals.

The experiments were conducted using Asteroid toolkit \cite{Pariente2020Asteroid}. The Beam-TasNet was composed of two modules, MC-Conv-TasNet and MVDR beamforming. Unlike \cite{Ochiai2020BeamTasNetTA}, we did not use voice activity detection-based refinement for simplicity and fair comparison. We trained two stages jointly using permutation invariant training (PIT) and an SNR loss \cite{Roux2019SDRH}. All models were trained with $4$-second segments and a maximum of $150$ epochs. The detailed model architecture is listed in Table \ref{tab:arch}, where the Beam-Guided TasNet had a roughly equal number of parameters with the baseline Beam-TasNet. Without iterations of the second-stage model, the proposed model has approximately the same computation cost as the baseline since they have a similar total number of parameters and the TasNet occupied most computation. One more iteration of the second stage model yields half the computation cost of the baseline model.

\begin{table}[]
	\centering
	\caption{The settings of the hyper-parameters of MC-Conv-TasNet in the baseline Beam-TasNet and the proposed Beam-Guided TasNet with 2 stages. The notations follow \cite{Luo2019ConvTasNetSI}.}
   \label{tab:arch}
   \resizebox{0.7\linewidth}{!}{
	\begin{tabular}{ccc}
	\toprule
	Hyper-parameter & Baseline  & First/Second stage  \\
	\midrule
	$N$            & $512$                  & $256$                 \\
	$L$            & $16$                   & $16$                 \\
	$B$            & $128$                  & $128$                 \\
	$S_c$          & $128$                  & $128$                 \\
	$H$            & $512$                  & $256$                 \\
	$P$            & $3$                    & $3$                 \\
	$X$            & $8$                    & $8$                 \\
	$R$            & $3$                    & $3$                 \\
	\midrule
	Model size (M)     & $5.4$ & $2.7$/$2.8$ \\
	\bottomrule
	\end{tabular}}
\end{table}

We used BSS-Eval SDR \cite{Vincent2006PerformanceMI} and WERs as the evaluation metrics. The SDR metric was calculated by comparing the estimated $\hat{x}_{s,1}$ or $\hat{z}_{s,1}$ with the reference signal $x_{s,1}$. The automatic speech recognition (ASR) system was trained following the scripts offered by the spatialized multi-speaker WSJ (SMS-WSJ) dataset \cite{Drude2019SMSWSJDP} to make the WER results reproducible. 

\section{Results and discussion}
\label{sec:res_dis}
\begin{table}[tbh]
	\centering
	\caption{Comparison of Beam-TasNet and Beam-Guided TasNet under the non-causal condition. The angle feature (AF) \cite{Gu2019NeuralSF} was obtained by the direction calculated by SRP-PHAT \cite{Brandstein1997ARM}. $\sharp$ means MC-Conv-TasNet is a MIMSO model.  $\dagger$ means the second-stage model is trained without unfolding.}
	\label{tab:exp2-stage}
	\resizebox{0.95\linewidth}{!}{
	\begin{tabular}{lclcccc}
	\toprule
	\multirow{2}{*}{Model} & Iteration & \multirow{2}{*}{Input} & \multicolumn{2}{c}{SDR$_\uparrow$ (dB)} & \multicolumn{2}{c}{WER$_\downarrow$ (\%)} \\
	  &  number  &      &     $\hat{z}_{s,1}$     &      $\hat{x}_{s,1}$ &  $\hat{z}_{s,1}$     &     $\hat{x}_{s,1}$           \\
	\midrule
	Beam-TasNet$^\sharp$  & -   & $y_{c}$ & $12.7$       		  & $17.2$ & $21.8$ & $14.0$ \\
	Beam-TasNet  & -   & $y_{c}$ & $12.7$       		  & $17.4$ & $22.1$ & $13.4$ \\
	\rowcolor{mygray}
	1-Stage  & -   & $y_{c}$   & 10.5 & 15.9 & 29.8 & 14.8 \\
	\midrule
	2-Stage  & 1  &  $y_{c}$ \& $\hat{z}_{s,c}^{(1)}$                  & $12.5$ & $17.1$ & $24.1$ & $14.6$ \\
	2-Stage  & 1   & $y_{c}$ \& AF                     			& $12.7$ & $17.5$ & $21.1$ & $13.6$ \\
	2-Stage  & 1   &  $y_{c}$ \& $\hat{x}_{s,c}^{(1)}$                  & 18.2 & $\mathbf{19.1}$ & $14.0$ & $\mathbf{12.3}$ \\
	2-Stage  & 1   &  $\hat{x}_{s,c}^{(1)}$                  & $17.4$ & $17.3$ & $14.0$ & $13.0$ \\
	\midrule
	3-Stage     & -            & $y_{c}$ \& $\hat{x}_{s,c}^{(2)}$  & $20.8$  & $19.7$ & $12.8$ & $12.3$ \\
	2-Stage$^\dagger$   & 2    & $y_{c}$ \& $\hat{x}_{s,c}^{(2:1)}$ & $19.7$ 	  & $19.7$ & $13.1$ & $12.2$ \\
	2-Stage  & 2     & $y_{c}$ \& $\hat{x}_{s,c}^{(2:1)}$ & $20.7$ 	  & $20.0$ & $12.9$ & $12.1$ \\
	2-Stage  & 4     & $y_{c}$ \& $\hat{x}_{s,c}^{(2:3)}$ & $\mathbf{21.5}$ & 20.3 & 12.8 & $\mathbf{12.1}$ \\
	\midrule
	Oracle IRM   & -    & - &  \multicolumn{1}{>{\columncolor{mygray}}c}{$12.9$} & $17.6$ & \multicolumn{1}{>{\columncolor{mygray}}c}{$12.4$} & $12.8$ \\
	Oracle signal & -    & - & \multicolumn{1}{>{\columncolor{mygray}}c}{$\infty$} & $\mathbf{23.5}$ & \multicolumn{1}{>{\columncolor{mygray}}c}{$\mathbf{11.7}$} & $11.9$ \\
	\bottomrule
	\end{tabular}}
\end{table}

\begin{table}[tbh]
	\caption{The performance of the causal systems. The gray cells share the same results with those in Table \ref{tab:exp2-stage}.}
	\label{tab:expcausal}
	\centering
	\resizebox{0.95\linewidth}{!}{
	\begin{tabular}{lccccccc}
	\toprule
	\multirow{2}{*}{Model} & Iteration & \multirow{2}{*}{Causal} & \multicolumn{2}{c}{SDR$_\uparrow$ (dB)} & \multicolumn{2}{c}{WER$_\downarrow$ (\%)} \\
							   & number &       &  $\hat{z}_{s,1}$     &     $\hat{x}_{s,1}$ &  $\hat{z}_{s,1}$     &     $\hat{x}_{s,1}$ \\
							   \midrule
	Beam-TasNet     &  -          & \cmark & $9.0$                 & $11.4$ & $33.6$ & $21.4$ \\
	\rowcolor{mygray} 
	1-Stage   & -    & \xmark  & $11.7$ & $16.7$ & $25.4$ & $14.0$ \\
	1-Stage    & -                 & \cmark & $8.6$        	& $10.9$             &  $35.1$  &  $22.7$ \\
	\midrule
	2-Stage   & 1           & \cmark & $13.1$       	& $12.2$             &  $19.7$  &  $20.0$ \\
	2-Stage  & 2   & \cmark & $13.9$       & $12.5$ & $18.7$ & $19.4$ \\
	2-Stage  & 4   & \cmark & $\mathbf{14.0}$       & $12.3$ & $\mathbf{18.6}$ & $19.4$ \\
	\midrule
	Oracle IRM & -  & \cmark &  \multicolumn{1}{>{\columncolor{mygray}}c}{$12.9$}  & $14.0$  & \multicolumn{1}{>{\columncolor{mygray}}c}{$12.4$} & $13.6$     \\
	Oracle signal & -  & \cmark & \multicolumn{1}{>{\columncolor{mygray}}c}{$\infty$} & $\mathbf{18.0}$ & \multicolumn{1}{>{\columncolor{mygray}}c}{$\mathbf{11.7}$} & $13.2$ \\
	\bottomrule
	\end{tabular}}
\end{table}

This section first performed an ablation study of Beam-Guided TasNet and compared the performance with the baseline Beam-TasNet and the oracle MVDR. Then, a causal framework was explored to illustrate the effectiveness of the framework without future information. Here we chose $n=4$ to obtain the best performance. Finally, we visualized the iterative processing to demonstrate how the framework boosts the performance with the guide of MVDR.

Table \ref{tab:exp2-stage} lists SDR and WER results on the baseline Beam-TasNet and the proposed Beam-Guided TasNet under the non-causal condition. The baseline Beam-TasNet achieved an SDR of $17.4$ dB, $0.6$ dB higher than \cite{Ochiai2020BeamTasNetTA}. We use MIMMO to directly generate signal for all channels. The performance on MVDR is sightly improved might due to the MIMMO considering the relation of channels. The first stage model adopted a small-sized model and achieved an SDR degradation of 1.5 dB and a WER degradation of 1.4\%. The second part of Table \ref{tab:exp2-stage} showed that using the second stage yielded SDR improvement and WER reduction with extra input of $\hat{z}_{s,c}$, angle feature and $\hat{x}_{s,c}$. The one with $\hat{x}_{s,c}$ obtained the best performance with the SDR improved by $3.2$ dB and the WER reduced by $2.5\%$ compared with the first stage. The MVDR beamformer is thought to play a crucial role in performance improvement since its output $\hat{x}_{s,c}^{(1)}$ presented a much higher SDR than $\hat{z}_{s,c}^{(1)}$. Only using $\hat{x}_{s,c}^{(1)}$ was found worse than using $y_{c}$ \& $\hat{x}_{s,c}^{(1)}$. In the third part of Table \ref{tab:exp2-stage}, the SDR and the WER was optimized to $19.7$ dB and $12.2\%$ with 2 iterations, which was worse than 3-stage deep unfolding model. However, using unfolding training loss (Eq.(\ref{eq:unfolding_loss})) could achieve similar performance. Ater 4 iterations, we got an SDR of $21.5$ dB and a WER of $12.1\%$, illustrating the effectiveness of shared parameters. On the other hand, for oracle MVDR, $\hat{z}_{s,1}$ equals to $x_{s,1}$ for the oracle signal. $\hat{z}_{s,1}$ was calculated based on the ideal ratio masks (IRMs) for oracle mask, which uses a window size of 32ms and a hop size of 16ms (Appendix C). The proposed Beam-Guided TasNet dramatically narrowed the SDR and the WER gap with the oracle signal-based MVDR to $2.0$ dB and $0.2\%$ and exceeded those of the oracle mask-based MVDR by $3.9$ dB and $0.3\%$, respectively.


Table \ref{tab:expcausal} lists results with the causal model. Introducing causality into MC-Conv-TasNet and MVDR degraded the performance. With the Beam-Guided TasNet and iterative processing, the SDR and the WER was optimized from $11.4$ dB and $21.4\%$ to $14.0$ dB and $18.6\%$. Again, the Beam-Guided TasNet exceeded that of the oracle mask-based MVDR and the baseline Beam-TasNet by $3.0$ dB and $2.6$ dB, respectively. 

\begin{figure}[t]
	\centering
    \resizebox{!}{0.65\linewidth}{
	\includegraphics[width=1.0\linewidth]{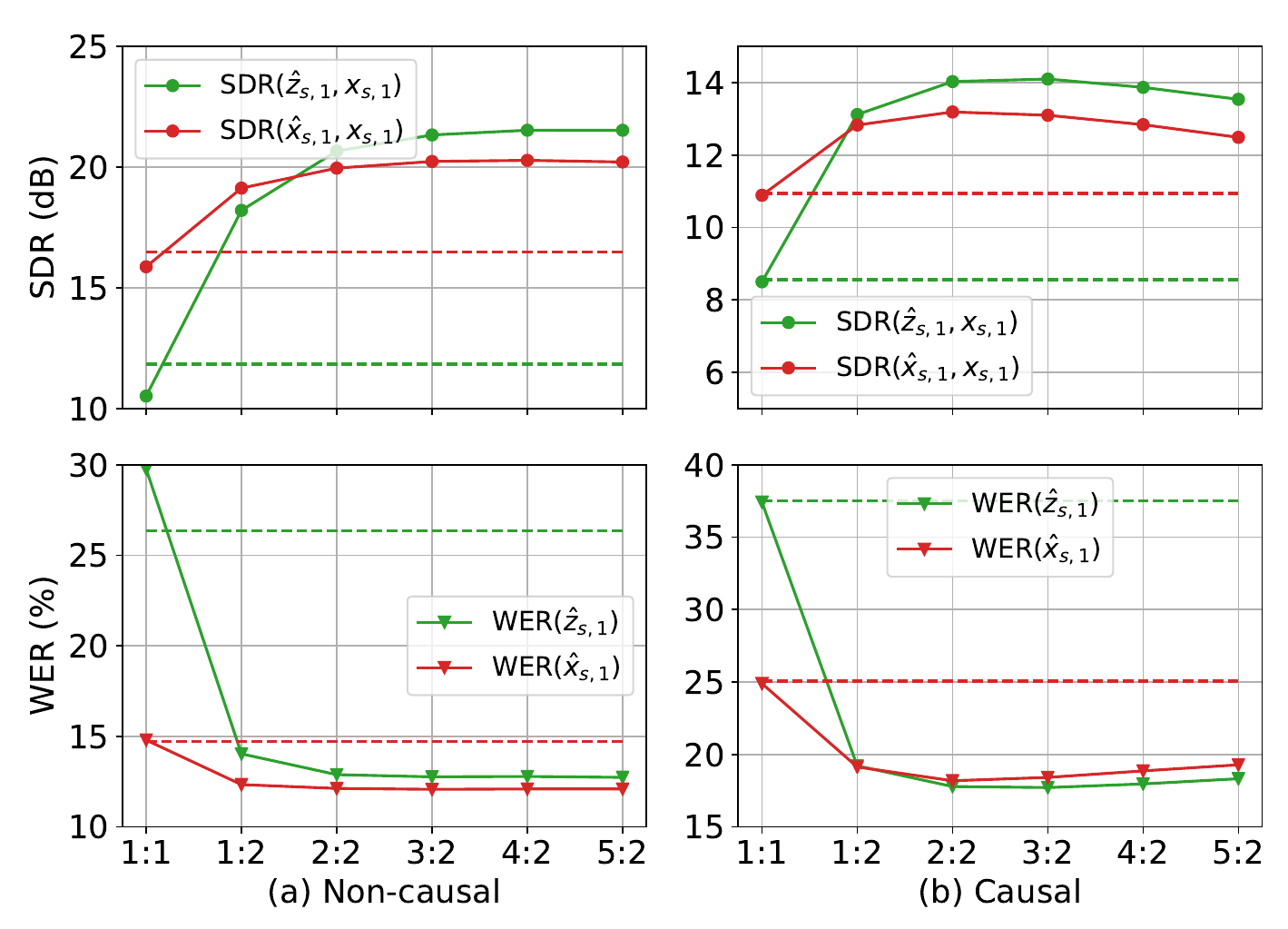} }
	\caption{SDR(dB)/WER(\%) vs. stage:iteration ($1/2:n$) under the causal/non-causal condition. The dashed lines are the results of the baseline Beam-TasNet.}
	\label{fig:iter_sdr2}
\end{figure}

The iterative processing is visualized in Fig.\ref{fig:iter_sdr2}, where the SDR and WER curves exhibit a nearly same trend on the non-causal and causal setting. We explain the following $3$ phenomena. First, the lines of SDRs raise and intersect, indicating that the Beam-Guided TasNet took the strength of MC-Conv-TasNet and MVDR to optimize each other. With a more accurate estimation of SCMs, the MVDR beamforming got closer to its upper bound gradually. However, the output of MC-TasNet in the current iteration could always achieve a better SDR than the output of MVDR in the previous iteration, which made $\hat{z}_{s,1}^{(2:n)}$ surpass $\hat{x}_{s,1}^{(2:n)}$ at some point. Second, we found that after 3 or 4 iterations, the Beam-Guided TasNet could achieve best performance. Third, the WER gap between $\hat{z}_{s,1}^{(2:n)}$ and $\hat{x}_{s,1}^{(2:n)}$ was eliminated after iterations. Under the non-causal condition, the distortionless $\hat{x}_{s,1}^{(2:n)}$ exhibited slightly lower WERs. Under the causal condition, however, the WER curve indicated that $\hat{z}_{s,1}^{(2:n)}$ obtained better signal quality due to the inaccurate MVDR filter.

We list experiment results on unmatched noisy condition, multi-speaker condition, learning anechoic signals and other model test on LibriCSS in Appendix D-G.

\section{Conclusion}

In this paper, we propose the Beam-Guided TasNet, which refines the multi-channel BSS iteratively with the guide of beamforming. The experiments presented considerable SDR improvement of $4.1$ dB and $2.6$ dB compared with the baseline Beam-TasNet under the non-causal and causal condition, respectively. In future work, we will further explore the design of MIMMO with novel network architectures.
\newpage

\section{Appendix}

\subsection{A. Frame-by-frame processing}
\label{append:causal_processing}

For online frame-by-frame processing, the permutation solver calculates metrics based on the received signal to conduct source reorder in a frame-by-frame method. In our practice, the distance measurement methods, such as Euclidean norm and correlation, can achieve similar performance. Here we use SNR to reorder the sources, which corresponds to Euclidean norm. The causal permutation solver obtains the order $\hat{\pi}_{c,t}$, which can be expressed as,
\begin{equation}
   \hat{\pi}_{c,t}=\underset{\pi_{c,t}}{\operatorname{argmax}} \sum_{s=1}^{S} \text{SNR}\left(\hat{x}_{s,1}[0:n_t], \hat{x}_{\pi_{c,t}(s),c}[0:n_t]\right),
\end{equation}
where $n_t$ denotes the number of received samples until frame $t$. The SCMs are updated as the followings:
\begin{equation}
	\mathbf{\hat{\Phi}}_{t,f}^{\mathrm{Target}_{s}}=\frac{t-1}{t}\mathbf{\hat{\Phi}}_{t-1,f}^{\mathrm{Target}_{s}}+\frac{1}{t}\hat{\mathbf{Z}}_{s, t, f} \hat{\mathbf{Z}}_{s, t, f}^{\mathrm{H}},
\end{equation}
\begin{align}
	\mathbf{\hat{\Phi}}_{t,f}^{\mathrm{Interfer}_{s}}=&\frac{t-1}{t}\mathbf{\hat{\Phi}}_{t-1,f}^{\mathrm{Interfer}_{s}} \\
	&+\frac{1}{t}(\mathbf{Y}_{t, f}-\hat{\mathbf{Z}}_{s, t, f})(\mathbf{Y}_{t, f}-\hat{\mathbf{Z}}_{s, t, f})^{\mathrm{H}},
\end{align}
where $\hat{\mathbf{Z}}_{s, t, f}$ is reordered by $\hat{\pi}_{c,t}$.

\subsection{B. Theoretical explanation}
Different from Beam-TasNet, the proposed iterative scheme focus on finding a distribution $p(y_c;x_{s,c})$ parameterized by $x_{s,c}$, which maximizes the probability of generating the observed data. According to \cite{DBLP:books/lib/Bishop07}, the loglikelihood $\log p(y_c;x_{s,c})$ can be decomposed into 2 terms using latent variable $\mathbf{\Phi}$:
\begin{align}
\label{eq:em}
	\log p\left(y_{c} ; x_{s,c}\right)=&\text{KL}\left[q(\mathbf{\Phi}) \| p\left(\mathbf{\Phi} \mid y_{c} ; x_{s,c}\right)\right]\\
	&+E_{q(\mathbf{\Phi})}\left[\log \frac{p\left(y_{c}, \mathbf{\Phi} ; x_{s,c}\right)}{q(\mathbf{\Phi})}\right]
\end{align}
where $\mathbf{\Phi}$ is the spatial correlation matrix $\mathbf{\hat{\Phi}}_{f}^{\mathrm{{Target}_{s}}}$ and $\mathbf{\hat{\Phi}}_{f}^{\mathrm{{Interfer}_{s}}}$. We can use MC-Conv-TasNet to estimate signals and then obtain $\mathbf{\hat{\Phi}}$ with the given $y_c$ and estimated parameters $\hat{x}_{s,c}$, which corresponds to $p(\mathbf{\Phi}|y_{c};x_{s,c})$. Since the neural network directly generates estimation, we can view the distribution as an impulse function. Then by setting $q(\Phi)=p(\Phi|y_{c};x_{s,c})$, maximizing the second item is equal to $\hat{x}_{c}=\operatorname{argmax}_{x_{s,c}} p\left(y_{c}, \hat{\mathbf{\Phi}};x_{s,c}\right)$. With Bayes’ rule, the optimal $\hat{x}_c$ is equal to $argmax_{x_c}p(x_c|y_c,\hat{\mathbf{\Phi}})$, which can be viewed as MVDR beamforming. Different from classic statistical models \cite{DBLP:conf/eusipco/ItoAN16,DBLP:journals/taslp/HiguchiIAYDN17}, TasNet does not guarantee the estimated $\mathbf{\hat{\Phi}}$ closer to the oracle one. Thus, the proposed method may exhibit performance degradation in iterations.

\subsection{C. The effect of STFT window size on MVDR}
\label{append:exp_windowsize}

\begin{table}[tbh]
	\caption{The performance of different window size for oracle IRMs.}
	\label{tab:exp_windowsize}
	\centering
	\resizebox{0.98\linewidth}{!}{
	\begin{tabular}{cccccc}
	\toprule
	Window size & \multirow{2}{*}{Causal} & \multicolumn{2}{c}{SDR$_\uparrow$ (dB)} & \multicolumn{2}{c}{WER$_\downarrow$ (\%)} \\
	(ms) &	&  $\hat{z}_{s,1}$     &     $\hat{x}_{s,1}$ &  $\hat{z}_{s,1}$     &     $\hat{x}_{s,1}$ \\
							   \midrule
	512 & \xmark &  11.0 & 14.5 & 28.1 & 15.8 \\
	32 & \xmark &  \textbf{12.9} & \textbf{17.6} & \textbf{12.4} & \textbf{12.8} \\
	512 & \cmark &  11.0 & 10.6 & 28.1 & 20.9 \\
	32 & \cmark &  \textbf{12.9} & \textbf{14.0} & \textbf{12.4} & \textbf{13.6} \\
	\bottomrule
	\end{tabular}}
\end{table}

The STFT settings affect the performance of oracle IRMs. A longer window size and stride will lead to worse SDRs as the phase plays a more important role. A window size of 512ms results in an SDR of 11.0dB (w/o MVDR) and 14.7dB (w/ MVDR), similar to the Beam-TasNet paper, while a window size of 32ms results in an SDR of 12.9dB (w/o MVDR) and 17.6dB (w/ MVDR).

\subsection{D. Unmatched noisy condition}
\label{append:unmatched_noisy_condition}

To evaluate the proposed framework under the noisy condition, we simulated the noisy training and evaluation sets by mixing WSJ0-2MIX dataset with real recorded noise from the REVERB challenge \cite{DBLP:journals/ejasp/KinoshitaDGHHKL16}. The training set contains noise recorded in a small room with an SNR range from 10 dB to 20 dB. The evaluation set contains noise recorded in a medium and a large room with an SNR range from 0 dB to 10 dB.

\begin{table}[tbh]
	\caption{The performance of non-causal models under the unmatched noisy condition.}
	\label{tab:exp_unmatched_noisy_condition}
	\centering
	\resizebox{0.98\linewidth}{!}{
	\begin{tabular}{lccccc}
	\toprule
	\multirow{2}{*}{Model} & Iteration & \multicolumn{2}{c}{SDR$_\uparrow$ (dB)} & \multicolumn{2}{c}{WER$_\downarrow$ (\%)} \\
					& number	&  $\hat{z}_{s,1}$     &     $\hat{x}_{s,1}$ &  $\hat{z}_{s,1}$     &     $\hat{x}_{s,1}$ \\
							   \midrule
	Beam-TasNet  & -    & $10.5$                 & $14.0$ & $31.1$ & $19.6$ \\
	1-Stage  & -     & $9.7$ & $13.6$ & $33.7$ & $20.3$ \\
	2-Stage & 1     & $15.2$             	& $15.1$             &  $17.7$  &  $18.1$ \\
	2-Stage & 2     & $16.2$       & $15.5$ & $16.5$ & $17.7$ \\
	2-Stage & 4     & $\mathbf{16.5}$       & $15.6$ & $\mathbf{16.3}$ & $17.7$ \\
	Oracle IRM   & -&  $12.3$ & $15.4$ & $12.6$ & $17.2$ \\
	Oracle signal & -& $\infty$ & $\mathbf{18.2}$ & $\mathbf{11.7}$ & $16.5$ \\ 
	\bottomrule
	\end{tabular}}
\end{table}

The experiment result in Table \ref{tab:exp_unmatched_noisy_condition} indicates that the proposed framework can deal with the noisy condition under unmatched noise settings. Compared with the baseline Beam-TasNet, our method achieved an SDR improvement of 2.5 dB and a WER reduction of 3.3\%.

\subsection{E. Multi-speaker condition}
\label{append:multi-speaker_condition}

We deployed the proposed methods on 2- and 3-speaker conditions with a non-causal model using the 2- and 3-speaker spatialized WSJ0-2MIX and WSJ0-3MIX datasets. We used A2PIT \cite{DBLP:conf/interspeech/KandaHTFNW19} for training, which can be integrated with the proposed Beam-guided TasNet by introducing multiple outputs.

\begin{table}[t]
	\caption{The performance on the 2-/3-speaker dataset using non-causal models.}
	\label{tab:multi-speaker_condition}
	\centering
	\resizebox{0.95\linewidth}{!}{
	\begin{tabular}{clccccc}
	\toprule
	Speaker & \multirow{2}{*}{Model}  & Iteration  & \multicolumn{2}{c}{SDR$_\uparrow$ (dB)} & \multicolumn{2}{c}{WER$_\downarrow$ (\%)} \\
	number & & number	&  $\hat{z}_{s,1}$     &     $\hat{x}_{s,1}$ &  $\hat{z}_{s,1}$     &     $\hat{x}_{s,1}$ \\
							   \midrule
	\multirow{5}{*}{2} & Beam-TasNet  & -    & $11.8$                 & $16.7$ & $25.3$ & $14.0$ \\
	& 1-Stage  & -     & $11.0$ & $16.1$ & $28.4$ & $14.6$ \\
	& 2-Stage & 1              & $18.4$       	& $19.1$             &  $14.2$  &  $12.4$ \\
	& 2-Stage & 2     & $20.0$       & $19.8$ & $13.2$ & $12.1$ \\
	& 2-Stage & 4     & $\mathbf{20.9}$       & $20.3$ & $13.1$ & $\mathbf{11.9}$ \\
	& Oracle IRM    & - & 12.9 & 17.6 & 12.4 & 12.8 \\
	& Oracle signal & - & $\infty$ & \textbf{23.5} & \textbf{11.7} & 11.9 \\
	\midrule
	\multirow{5}{*}{3} & Beam-TasNet  & -    & $7.3$                 & $11.4$ & $48.0$ & $23.8$ \\
	& 1-Stage  & -     & $6.4$ & $10.6$ & $52.6$ & $25.8$ \\
	& 2-Stage & 1              & $12.4$       	& $13.7$             &  $22.5$  &  $17.2$ \\
	& 2-Stage & 2     & $14.6$       & $14.8$ & $17.5$ & $15.5$ \\
	& 2-Stage & 4     & $\mathbf{15.8}$       & $15.5$ & $15.9$ & $\mathbf{14.5}$ \\
	& Oracle IRM  & - & 9.8 & 14.8 & 12.2 & 14.8 \\ 
	& Oracle signal & - & $\infty$ & \textbf{22.1} & \textbf{11.7} & 12.3 \\
	\bottomrule
	\end{tabular}}
\end{table}

The experimental results are listed in Table \ref{tab:multi-speaker_condition}. We have found that the proposed Beam-guided TasNet could outperform the Beam-TasNet consistently under the 2- and 3-speaker condition.

\subsection{F. Learning anechoic signals}

Previous experiments use models to learn single-speaker reverberant signals. Here we set the learning target to single-speaker anechoic signals to perform both dereverberation and separation tasks. 

\begin{table}[h]
	\caption{The performance of non-causal models on spatialized WSJ0-2MIX. The learning target and the reference signal for SDR calculation is single-speaker anechoic signals.}
	\label{tab:learn_anechoic}
	\centering
	\resizebox{0.98\linewidth}{!}{
	\begin{tabular}{lccccc}
	\toprule
	\multirow{2}{*}{Model} & Iteration & \multicolumn{2}{c}{SDR$_\uparrow$ (dB)} & \multicolumn{2}{c}{WER$_\downarrow$ (\%)} \\
	& number	&  $\hat{z}_{s,1}$     &     $\hat{x}_{s,1}$ &  $\hat{z}_{s,1}$     &     $\hat{x}_{s,1}$ \\
	\midrule
	Beam-TasNet  & -    & 10.8  & 14.6 	& 29.8 	& 15.2 \\
	1-Stage      & -    & 9.4  & 14.0 	& 38.9 	& 17.1 \\
	2-Stage      & 1    & 14.5  & 16.4 	& 17.9 	& 13.6 \\
	2-Stage      & 2    & 16.5  & 17.1 	& 14.9 	& 12.8 \\
	2-Stage      & 4    & 17.1  & \textbf{17.3} 	& 14.2 	& \textbf{12.4} \\
	Oracle IRM                     &  -    & 11.4 & 12.0  & 11.1 & 15.5 \\
	Oracle signal                 & -    & $\mathbf{\infty}$ & \textbf{21.1} & \textbf{10.2} & 11.4 \\
	\bottomrule
	\end{tabular}}
\end{table}

The experiment results in Table \ref{tab:exp_learning_anechoic} exhibit that the Beam-guided TasNet achieves an SDR of 17.3dB and a WER of 12.4\%, far exceeding Beam-TasNet.

\subsection{G. Experiments on LibriCSS with frequency-domain model}

\begin{table}[h]
	\caption{The performance of non-causal models on LibriCSS.}
	\label{tab:exp_learning_anechoic}
	\centering
	\resizebox{0.98\linewidth}{!}{
	\begin{tabular}{lccccccc}
	\toprule
	\multirow{2}{*}{Model} & Iteration &  \multicolumn{6}{c}{WER$_\downarrow$ (\%)} \\
	& number	&  0S     &     0L &  OV10     &  OV20     & OV30     &   OV40 \\
	\midrule
	Unprocessed & -     & 11.8 & 11.7 & 18.8 & 27.2 & 35.6 & 43.3 \\
	DPT-FSNET    & -    & 7.1  & 7.3	& 7.6 	& 8.9 &  10.8 & 11.3 \\
	1-Stage      & -    &  7.3 & 7.3 & 7.8 & 8.9 & 10.6 & 11.1 \\
	2-Stage      & 1    &  7.1 & 7.1 & 7.1 & 8.0 & 9.2  & 9.7  \\
	2-Stage      & 2    &  \textbf{7.0} & \textbf{7.1} & \textbf{6.9} & 7.9 & \textbf{8.8}  & 9.3  \\
	2-Stage      & 4    &  \textbf{7.0} & \textbf{7.1} & 7.0 & \textbf{7.7} & \textbf{8.8}  & \textbf{9.0}  \\
	\bottomrule
	\end{tabular}}
\end{table}

LibriCSS is a real-recorded dataset. The ASR engine uses the original hybrid model \cite{DBLP:conf/icassp/ChenYLZMLWXL20}. We validate the iterative framework on a frequency-domain model, named DPT-FSNET\cite{dangfeng}. After iterations, we achieve a WER of 9.0\% on OV40 subset, 3.0\% lower than DPT-FSNETs.
\bibliographystyle{IEEEtran}

\bibliography{mybib}

\begin{thebibliography}{10}
\providecommand{\url}[1]{#1}
\csname url@samestyle\endcsname
\providecommand{\newblock}{\relax}
\providecommand{\bibinfo}[2]{#2}
\providecommand{\BIBentrySTDinterwordspacing}{\spaceskip=0pt\relax}
\providecommand{\BIBentryALTinterwordstretchfactor}{4}
\providecommand{\BIBentryALTinterwordspacing}{\spaceskip=\fontdimen2\font plus
\BIBentryALTinterwordstretchfactor\fontdimen3\font minus
  \fontdimen4\font\relax}
\providecommand{\BIBforeignlanguage}[2]{{%
\expandafter\ifx\csname l@#1\endcsname\relax
\typeout{** WARNING: IEEEtran.bst: No hyphenation pattern has been}%
\typeout{** loaded for the language `#1'. Using the pattern for}%
\typeout{** the default language instead.}%
\else
\language=\csname l@#1\endcsname
\fi
#2}}
\providecommand{\BIBdecl}{\relax}
\BIBdecl

\bibitem{Erdogan2016ImprovedMB}
H.~Erdogan, J.~R. Hershey, S.~Watanabe, M.~I. Mandel, and J.~L. Roux,
  ``Improved {MVDR} beamforming using single-channel mask prediction
  networks,'' in \emph{Interspeech 2016, 17th Annual Conference of the
  International Speech Communication Association, San Francisco, CA, USA,
  September 8-12, 2016}, N.~Morgan, Ed.\hskip 1em plus 0.5em minus 0.4em\relax
  {ISCA}, 2016, pp. 1981--1985.

\bibitem{Kanda2019GuidedSS}
N.~Kanda, C.~B{\"o}ddeker, J.~Heitkaemper, Y.~Fujita, S.~Horiguchi,
  K.~Nagamatsu, and R.~Haeb-Umbach, ``Guided source separation meets a strong
  asr backend: Hitachi/paderborn university joint investigation for dinner
  party asr,'' in \emph{INTERSPEECH}, 2019.

\bibitem{Luo2019ConvTasNetSI}
Y.~Luo and N.~Mesgarani, ``Conv-tasnet: Surpassing ideal time-frequency
  magnitude masking for speech separation,'' \emph{{IEEE} {ACM} Trans. Audio
  Speech Lang. Process.}, vol.~27, no.~8, pp. 1256--1266, 2019.

\bibitem{Ochiai2020BeamTasNetTA}
T.~Ochiai, M.~Delcroix, R.~Ikeshita, K.~Kinoshita, T.~Nakatani, and S.~Araki,
  ``Beam-tasnet: Time-domain audio separation network meets frequency-domain
  beamformer,'' in \emph{2020 {IEEE} International Conference on Acoustics,
  Speech and Signal Processing, {ICASSP} 2020, Barcelona, Spain, May 4-8,
  2020}.\hskip 1em plus 0.5em minus 0.4em\relax {IEEE}, 2020, pp. 6384--6388.

\bibitem{Wang2018MultiChannelDC}
Z.~Wang, J.~L. Roux, and J.~R. Hershey, ``Multi-channel deep clustering:
  Discriminative spectral and spatial embeddings for speaker-independent speech
  separation,'' in \emph{2018 {IEEE} International Conference on Acoustics,
  Speech and Signal Processing, {ICASSP} 2018, Calgary, AB, Canada, April
  15-20, 2018}.\hskip 1em plus 0.5em minus 0.4em\relax {IEEE}, 2018, pp. 1--5.

\bibitem{Gu2019EndtoEndMS}
R.~Gu, J.~Wu, S.~Zhang, L.~Chen, Y.~Xu, M.~Yu, D.~Su, Y.~Zou, and D.~Yu,
  ``End-to-end multi-channel speech separation,'' \emph{CoRR}, vol.
  abs/1905.06286, 2019.

\bibitem{Souden2010OnOF}
M.~Souden, J.~Benesty, and S.~Affes, ``On optimal frequency-domain multichannel
  linear filtering for noise reduction,'' \emph{{IEEE} Trans. Speech Audio
  Process.}, vol.~18, no.~2, pp. 260--276, 2010.

\bibitem{molkov2019SpeakerBeamSA}
K.~Zmol{\'{\i}}kov{\'{a}}, M.~Delcroix, K.~Kinoshita, T.~Ochiai, T.~Nakatani,
  L.~Burget, and J.~Cernock{\'{y}}, ``Speakerbeam: Speaker aware neural network
  for target speaker extraction in speech mixtures,'' \emph{{IEEE} J. Sel. Top.
  Signal Process.}, vol.~13, no.~4, pp. 800--814, 2019.

\bibitem{Gu2019NeuralSF}
R.~Gu, L.~Chen, S.~Zhang, J.~Zheng, Y.~Xu, M.~Yu, D.~Su, Y.~Zou, and D.~Yu,
  ``Neural spatial filter: Target speaker speech separation assisted with
  directional information,'' in \emph{Interspeech 2019, 20th Annual Conference
  of the International Speech Communication Association, Graz, Austria, 15-19
  September 2019}, G.~Kubin and Z.~Kacic, Eds.\hskip 1em plus 0.5em minus
  0.4em\relax {ISCA}, 2019, pp. 4290--4294.

\bibitem{Gu2020TemporalSpatialNF}
R.~Gu and Y.~Zou, ``Temporal-spatial neural filter: Direction informed
  end-to-end multi-channel target speech separation,'' \emph{CoRR}, vol.
  abs/2001.00391, 2020.

\bibitem{Sonning2020PerformanceSO}
S.~Sonning, C.~Sch{\"{u}}ldt, H.~Erdogan, and S.~Wisdom, ``Performance study of
  a convolutional time-domain audio separation network for real-time speech
  denoising,'' in \emph{2020 {IEEE} International Conference on Acoustics,
  Speech and Signal Processing, {ICASSP} 2020, Barcelona, Spain, May 4-8,
  2020}.\hskip 1em plus 0.5em minus 0.4em\relax {IEEE}, 2020, pp. 831--835.

\bibitem{DBLP:journals/corr/HersheyRW14}
J.~R. Hershey, J.~L. Roux, and F.~Weninger, ``Deep unfolding: Model-based
  inspiration of novel deep architectures,'' \emph{CoRR}, vol. abs/1409.2574,
  2014.

\bibitem{DBLP:journals/taslp/WangWW21}
\BIBentryALTinterwordspacing
Z.~Wang, P.~Wang, and D.~Wang, ``Multi-microphone complex spectral mapping for
  utterance-wise and continuous speech separation,'' \emph{{IEEE} {ACM} Trans.
  Audio Speech Lang. Process.}, vol.~29, pp. 2001--2014, 2021. [Online].
  Available: \url{https://doi.org/10.1109/TASLP.2021.3083405}
\BIBentrySTDinterwordspacing

\bibitem{DBLP:conf/apsipa/TogamiMKYK20}
M.~Togami, Y.~Masuyama, T.~Komatsu, K.~Yoshii, and T.~Kawahara,
  ``Computer-resource-aware deep speech separation with a run-time-specified
  number of {BLSTM} layers,'' in \emph{Asia-Pacific Signal and Information
  Processing Association Annual Summit and Conference, {APSIPA} 2020, Auckland,
  New Zealand, December 7-10, 2020}.\hskip 1em plus 0.5em minus 0.4em\relax
  {IEEE}, 2020, pp. 788--793.

\bibitem{WichernAFZMCMR19}
G.~Wichern, J.~Antognini, M.~Flynn, L.~R. Zhu, E.~McQuinn, D.~Crow, E.~Manilow,
  and J.~L. Roux, ``Wham!: Extending speech separation to noisy environments,''
  in \emph{Interspeech 2019, 20th Annual Conference of the International Speech
  Communication Association, Graz, Austria, 15-19 September 2019}, G.~Kubin and
  Z.~Kacic, Eds.\hskip 1em plus 0.5em minus 0.4em\relax {ISCA}, 2019, pp.
  1368--1372.

\bibitem{Pariente2020Asteroid}
M.~Pariente, S.~Cornell, J.~Cosentino, S.~Sivasankaran, E.~Tzinis,
  J.~Heitkaemper, M.~Olvera, F.-R. Stöter, M.~Hu, J.~M. Martín-Doñas,
  D.~Ditter, A.~Frank, A.~Deleforge, and E.~Vincent, ``Asteroid: the
  {PyTorch}-based audio source separation toolkit for researchers,'' in
  \emph{Proc. Interspeech}, 2020.

\bibitem{Roux2019SDRH}
J.~L. Roux, S.~Wisdom, H.~Erdogan, and J.~R. Hershey, ``{SDR} - half-baked or
  well done?'' in \emph{{IEEE} International Conference on Acoustics, Speech
  and Signal Processing, {ICASSP} 2019, Brighton, United Kingdom, May 12-17,
  2019}.\hskip 1em plus 0.5em minus 0.4em\relax {IEEE}, 2019, pp. 626--630.

\bibitem{Vincent2006PerformanceMI}
E.~Vincent, R.~Gribonval, and C.~F{\'{e}}votte, ``Performance measurement in
  blind audio source separation,'' \emph{{IEEE} Trans. Speech Audio Process.},
  vol.~14, no.~4, pp. 1462--1469, 2006.

\bibitem{Drude2019SMSWSJDP}
L.~Drude, J.~Heitkaemper, C.~B{\"{o}}ddeker, and R.~Haeb{-}Umbach, ``{SMS-WSJ:}
  database, performance measures, and baseline recipe for multi-channel source
  separation and recognition,'' \emph{CoRR}, vol. abs/1910.13934, 2019.

\bibitem{Brandstein1997ARM}
M.~S. Brandstein and H.~F. Silverman, ``A robust method for speech signal
  time-delay estimation in reverberant rooms,'' in \emph{1997 {IEEE}
  International Conference on Acoustics, Speech, and Signal Processing,
  {ICASSP} '97, Munich, Germany, April 21-24, 1997}.\hskip 1em plus 0.5em minus
  0.4em\relax {IEEE} Computer Society, 1997, pp. 375--378.

\bibitem{DBLP:books/lib/Bishop07}
C.~M. Bishop, \emph{Pattern recognition and machine learning, 5th Edition: 10.1
  Variational Inference}, ser. Information science and statistics.\hskip 1em
  plus 0.5em minus 0.4em\relax Springer, 2007.

\bibitem{DBLP:conf/eusipco/ItoAN16}
N.~Ito, S.~Araki, and T.~Nakatani, ``Complex angular central gaussian mixture
  model for directional statistics in mask-based microphone array signal
  processing,'' in \emph{24th European Signal Processing Conference, {EUSIPCO}
  2016, Budapest, Hungary, August 29 - September 2, 2016}.\hskip 1em plus 0.5em
  minus 0.4em\relax {IEEE}, 2016, pp. 1153--1157.

\bibitem{DBLP:journals/taslp/HiguchiIAYDN17}
T.~Higuchi, N.~Ito, S.~Araki, T.~Yoshioka, M.~Delcroix, and T.~Nakatani,
  ``Online {MVDR} beamformer based on complex gaussian mixture model with
  spatial prior for noise robust {ASR},'' \emph{{IEEE} {ACM} Trans. Audio
  Speech Lang. Process.}, vol.~25, no.~4, pp. 780--793, 2017.

\bibitem{DBLP:journals/ejasp/KinoshitaDGHHKL16}
K.~Kinoshita, M.~Delcroix, S.~Gannot, E.~A.~P. Habets, R.~Haeb{-}Umbach,
  W.~Kellermann, V.~Leutnant, R.~Maas, T.~Nakatani, B.~Raj, A.~Sehr, and
  T.~Yoshioka, ``A summary of the {REVERB} challenge: state-of-the-art and
  remaining challenges in reverberant speech processing research,''
  \emph{{EURASIP} J. Adv. Signal Process.}, vol. 2016, p.~7, 2016.

\bibitem{DBLP:conf/interspeech/KandaHTFNW19}
N.~Kanda, S.~Horiguchi, R.~Takashima, Y.~Fujita, K.~Nagamatsu, and S.~Watanabe,
  ``Auxiliary interference speaker loss for target-speaker speech
  recognition,'' in \emph{Interspeech 2019, 20th Annual Conference of the
  International Speech Communication Association, Graz, Austria, 15-19
  September 2019}, G.~Kubin and Z.~Kacic, Eds.\hskip 1em plus 0.5em minus
  0.4em\relax {ISCA}, 2019, pp. 236--240.

\bibitem{DBLP:conf/icassp/ChenYLZMLWXL20}
\BIBentryALTinterwordspacing
Z.~Chen, T.~Yoshioka, L.~Lu, T.~Zhou, Z.~Meng, Y.~Luo, J.~Wu, X.~Xiao, and
  J.~Li, ``Continuous speech separation: Dataset and analysis,'' in \emph{2020
  {IEEE} International Conference on Acoustics, Speech and Signal Processing,
  {ICASSP} 2020, Barcelona, Spain, May 4-8, 2020}.\hskip 1em plus 0.5em minus
  0.4em\relax {IEEE}, 2020, pp. 7284--7288. [Online]. Available:
  \url{https://doi.org/10.1109/ICASSP40776.2020.9053426}
\BIBentrySTDinterwordspacing

\bibitem{dangfeng}
F.~Dang, H.~Chen, and P.~Zhang, ``Dpt-fsnet: Dual-path transformer based
  full-band and sub-band fusion network for speech enhancement,'' in \emph{2022
  {IEEE} International Conference on Acoustics, Speech and Signal Processing,
  {ICASSP} 2022}.\hskip 1em plus 0.5em minus 0.4em\relax {IEEE}, 2022.

\end{thebibliography}

\end{document}